\documentclass[11pt,a4paper]{article}

\usepackage{amsthm}
\usepackage{amsmath}
\usepackage{natbib}
\usepackage[colorlinks,citecolor=blue,urlcolor=blue,filecolor=blue,backref=page]{hyperref}
\usepackage{graphicx}
\usepackage{enumitem}
\usepackage{verbatim}
\usepackage{graphics}
\usepackage{color}
\usepackage{caption}
\usepackage{comment}
\usepackage{amssymb}
\usepackage{subcaption}
\usepackage{tikz}
\usepackage{algorithm}
\usepackage{multirow}

\title{Modeling goal chances in soccer: a Bayesian inference approach}
\author{Gavin A.~Whitaker$^{1,2}$\thanks{email: \texttt{gavin.whitaker@ucl.ac.uk}} \and\ Ricardo Silva$^{1,3}$ \and\ Daniel Edwards$^2$}
\date{\small $^{1}$Department of Statistical Science, University College London, London, WC1E 6BT \\
$^{2}$Stratagem Technologies, 19 Eastbourne Terrace, London, W2 6LG \\
$^{3}$The Alan Turing Institute, 96 Euston Road, London, NW1 2DB}

\begin{document}
\maketitle
\begin{abstract}
We consider the task of determining the number of chances a soccer 
team creates, along with the composite nature of each chance---the 
players involved and the locations on the pitch of the assist and the 
chance. We propose an interpretable Bayesian inference approach and 
implement a Poisson model to capture chance occurrences, from which we 
infer team abilities. We then use a Gaussian mixture model to capture 
the areas on the pitch a player makes an assist/takes a chance. This 
approach allows the visualization of differences between players in 
the way they approach attacking play (making assists/taking chances). 
We apply the resulting scheme to the 2016/2017 English Premier League, 
capturing team abilities to create chances, before highlighting key 
areas where players have most impact.
\end{abstract}


\section{Introduction}

Within this paper we look to explain an English Premier League team's 
style of attacking play; determining the number of chances a team 
creates, along with identifying the players involved and from where on 
the pitch the chance took place. 

The Premier League is an annual soccer league established in 1992 and 
is the most watched soccer league in the world \citep{yueh_2014, curley_2016}. 
It consists of 20 teams, who, over the course of a season, play every 
other team twice (both home and away), giving a total of 380 fixtures. 
It is the top division of English soccer, and every year 
the bottom 3 teams are relegated to be replaced by 3 teams from the 
next division down (the Championship). In recent times the Premier 
League has also become known as the richest league in the world 
\citep{deloitte_2016}, through both foreign investment and a 
lucrative deal for television rights \citep{rumsby_2016, bbc_2015}. 
To compete in the Premier League, teams employ different styles of 
play, often determined by the manager's personal preferences and the 
players who make up the team. Examples of attacking styles of play 
include counter attacking (quickly moving the ball into scoring range) 
or passing-build-up (many short passes to find a weakness in the 
oppositions defense). For further discussion of styles of play, we 
direct the reader to \citep{wendichansky_2016, huddleston_2018}. 

Methods to model a soccer team's style of play/behavior have been 
explored previously by a number of authors. \cite{lucey_2013} use 
occupancy maps defined using a given metric, for example, the mean or an 
entropy measure, to determine a team's style of play with the aim of 
showing that a team will aim to ``win home games and draw away ones.'' 
Occupancy maps are also used by \cite{bialkowski_2014}, who take 
spatio-temporal player tracking data and develop a method to automatically 
detect formation and player roles. \cite{bojinov_2016} utilize Gaussian 
processes to form a spatial map to capture each team's defensive 
strengths and weaknesses. \cite{pena_2012, pena_2014, pena_2015} employ 
methods from the Network analysis toolbox to draw conclusions about a 
team/player's use of possession. How the player's on a team interact 
is discussed in \cite{grund_2012}, and \cite{kim_2010} estimate the 
global movements of all players to predict short term evolutions of play. 
Outside of soccer, \cite{miller_2014} 
investigate shot selection amongst basketball players in the NBA, 
combining matrix factorization techniques with an intensity surface, 
modeled using a log-Gaussian Cox process. Defensive play in basketball 
is captured by \cite{franks_2015}, who take player tracking data and apply 
spatio-temporal processes, matrix factorization techniques and 
hierarchical regression models. 
 
More generally, the statistical modeling of sports has become a topic of increasing 
interest in recent times, as more data is collected on the sports we 
love, coupled with a heightened interest in the outcome of these 
sports, that is, the continuous rise of online betting. Soccer is 
providing an area of rich research, with the ability to capture the 
goals scored in a match being of particular interest, see 
\citep{dixon_1997, karlis_2003, baio_2010}. 
A player performance rating system (the EA Sports Player Performance 
Index) was developed by \cite{mchale_2012}, which aims to represent a 
player's worth in a single number, whilst \cite{mchale_2014} identify 
the goal scoring ability of players. \cite{whitaker_2017} rate players 
for a number of abilities, before using them to aid the prediction of 
goals scored. Finally \cite{kharrat_2017} develop a plus-minus rating 
system for soccer. 

In this paper we propose a method to capture the 
number of chances a team creates during a given section of a match, 
along with determining the players involved in a chance, where on the 
pitch the chance was created and where it was taken from. Our work 
differs from previous studies in this area in a number of ways. Firstly, 
previous work has used complete touch data (where every location that 
a player touches the ball in a game is recorded), to model a team's 
attacking play. Here, we use only the location of the assist and the 
chance. Thus, our proposed method is less computationally intensive 
and allows inferences from coarser and significantly cheaper data. 
Previous work has also focused on modeling the spatial dynamics of a 
team as a whole, whereas our method identifies the individual spatial 
contributions of players. Where specific players have been modeled in 
the past, this is often not accompanied by spatial analysis, instead 
player-to-player relationships are considered. We note that the model 
proposed within this paper has a wide variety of applications, of which 
we illustrate a few. 

The remainder of this article is organized as follows. The data is 
presented in Section~\ref{data}. In Section~\ref{model} we outline our 
model to capture a teams chances, before discussing an approach to 
identify the players involved with each chance and from which spatial 
locations. Applications are considered in Section~\ref{app} and a 
discussion is provided in Section~\ref{disc}.

\section{The data} \label{data}

The data available to us is Stratagem Technologies' Analyst data. This is a 
collection of data which marks the significant events during a soccer 
match; including goals, cards (both yellow and red) and chances created. 
For each of these events a time is recorded (in minutes), the team and player 
involved with the event, and for the goals/chances the location on the pitch 
is marked. If the event is a goal/chance, both the player taking the chance 
and the player assisting the chance are recorded (along with the spatial 
location of the chance and the assist). From here on in, we consider goals 
and chances to be the same for our purposes (a goal being a chance which 
is scored after all)---we refer to them collectively as chance. A section 
of the data is shown in Table~\ref{tab-data}. The data covers the 
2016/2017 English Premier League season and consists of roughly 32K 
events in total, which equates to approximately 85 events for each 
fixture in the dataset. We also have the date of each fixture.      

{
\begin{table} 
\centering
\footnotesize
\begin{tabular}{ccccccccccc}
\hline
\multirow{2}{*}{fixture} &	\multirow{2}{*}{date} & \multirow{2}{*}{team} &	\multirow{2}{*}{time} & \multirow{2}{*}{type} & event & assist & assist & assist & chance & chance \\
&&&&& player & player & x & y & x & y \\
\hline
2241765 & 2016-08-13 & 725 & 82.35 & Yellow card & 94174 & --- & --- & --- & --- & --- \\ 
2241765 & 2016-08-13 & 725 & 81.38 & Chance & 38569 & 38569 & -108 & 21 & -98 & 34 \\ 
2241765 & 2016-08-13 & 682 & 75.65 & Chance & 5724 & 11180 & 136 & 41 & 26 & 45 \\
2241765 & 2016-08-13 & 682 & 72.48 & Chance & 156662 & 159732 & 47 & 76 & 48 & 39 \\
\hline
\end{tabular}
\caption{A section of Stratagem Technologies' analyst data} \label{tab-data}
\end{table} 
}

\begin{figure}[t!]
      \centering
      \includegraphics[scale=0.48]{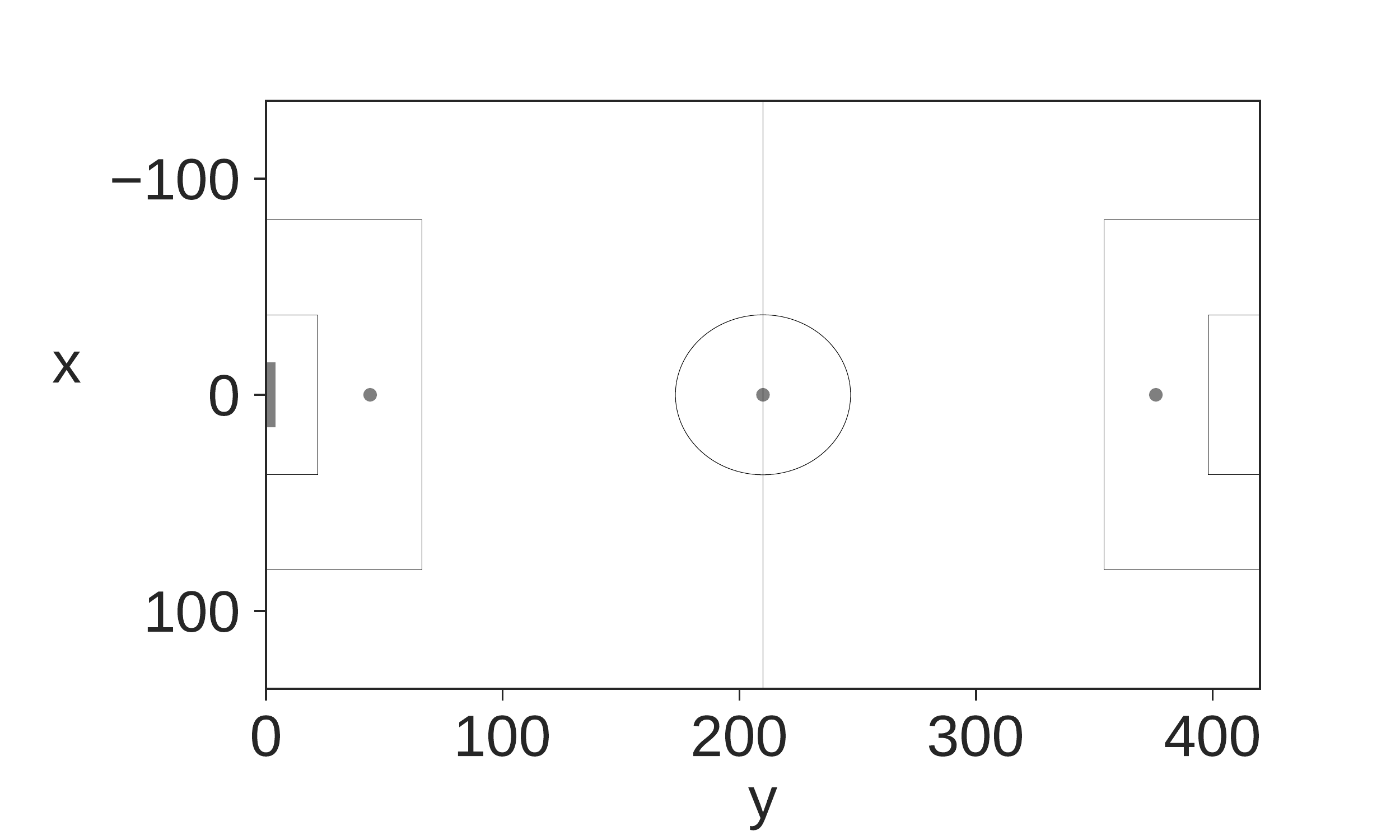}
      \caption{Map of the pitch, the point (0,0) represents the center of the defended goal (shaded box). Further key reference points are detailed in Table~\ref{tab-points}} \label{pitchfig}
\end{figure} 

\begin{table}[h!] 
\centering
\begin{tabular}{lcc}
\hline
Point &	$x$ & $y$ \\
\hline
Center of defended goal & 0 & 0  \\ 
Right goalpost & 15 & 0  \\ 
Left goalpost & -15 & 0  \\ 
6-yard box, right corner & 37 & 22 \\
6-yard box, left corner & -37 & 22 \\
Penalty spot & 0 & 44 \\
18-yard box, right corner & 81 & 66 \\
18-yard box, left corner & -81 & 66 \\
Center spot & 0 & 210 \\
\hline
\end{tabular}
\caption{Key reference points} \label{tab-points}
\end{table} 

Locations on the pitch are represented by $(x,y)$-coordinates with the 
$x$-axis running between the two touch-lines (width of the pitch) and 
the $y$-axis representing the length of the pitch between the goalposts. 
The spatial location is always recorded from the perspective of the 
attacking team, meaning the coordinate system does not need to be 
rotated to account for the second team, or to accommodate the fact that 
teams switch ends at half-time. The point $(0,0)$ marks the center of 
the defended goal, with the width of the pitch going from -136 to 136 
(left to right), and the pitch length running from 0 to 420. Explicitly, 
$x\in[-136,136]$ and $y\in[0,420]$. A map of the pitch is shown in 
Figure~\ref{pitchfig}, with some key reference points given in 
Table~\ref{tab-points}.

Further to the above, it is possible to extract additional statistics 
from the dataset. These include the game state and the red card state 
for a team at a given time point. The game state is the number of goals 
a team is winning or losing by at that point in time, for example, a 
team winning 1-0 would have a game state of +1, a team losing 1-3 would 
be -2, and, if the game is currently a draw, both teams would have a 
game state of 0. The red card state is defined similarly, and is the 
difference in the number of players on each team. To elucidate, if a 
team has a player sent off their red card state would be -1, whilst 
the opposition would be +1.

\section{The model} \label{model}

\begin{figure}
\centering
\begin{tikzpicture}[scale=7,>=latex]
     \node[draw,circle]
         (N) at (0.9,1.2) {\large $N$};
             
     \node[draw,circle]
         (e1) at (0.4,1) {\large $E_1$};
         
     \node[draw,circle]
         (a1) at (0.1,0.85) {\large $A_1$};
     \node[draw,circle]
         (c1) at (0.1,0.65) {\large $C_1$};
     \node[draw,circle]
         (xa1) at (0.35,0.65) {\large $\left(x_1^a,y_1^a\right)$};
     \node[draw,circle]
         (xd1) at (0.65,0.65) {\large $\left(\Delta_1^x,\Delta_1^y\right)$};

     \node at (0.55, 1) {{$\ldots$}};              
     \node at (0.9, 1) {{$\ldots$}}; 
     \node at (1.25, 1) {{$\ldots$}}; 
     
     \node (e2) at (0.7, 1.03) {{}};
     \node (e3) at (1.1, 1.03) {{}};
     
     \node[draw,circle]
         (eN) at (1.4,1) {\large $E_N$};
         
     \node[draw,circle]
         (aN) at (1.75,0.85) {\large $A_N$};
     \node[draw,circle]
         (cN) at (1.75,0.65) {\large $C_N$};
     \node[draw,circle]
         (xaN) at (1.48,0.65) {\large $\left(x_N^a,y_N^a\right)$};
     \node[draw,circle]
         (xdN) at (1.15,0.65) {\large $\left(\Delta_N^x,\Delta_N^y\right)$};

      
      \draw (e1) edge[->]  (N) ;
      \draw (e2) edge[->, black!50!]  (N) ;
      \draw (e3) edge[->, black!50!]  (N) ;
      \draw (eN) edge[->]  (N) ;
        
      \draw (a1) edge[->]  (e1) ;
      \draw (c1) edge[->]  (e1) ;      
      \draw (xa1) edge[->]  (e1) ;  
      \draw (xd1) edge[->]  (e1) ;   

      \draw (aN) edge[->]  (eN) ;
      \draw (cN) edge[->]  (eN) ;      
      \draw (xaN) edge[->]  (eN) ;  
      \draw (xdN) edge[->]  (eN) ;  

\end{tikzpicture}
\caption{Visual representation of the model for a single team in a given fixture} \label{pic-modelrep}
\end{figure}
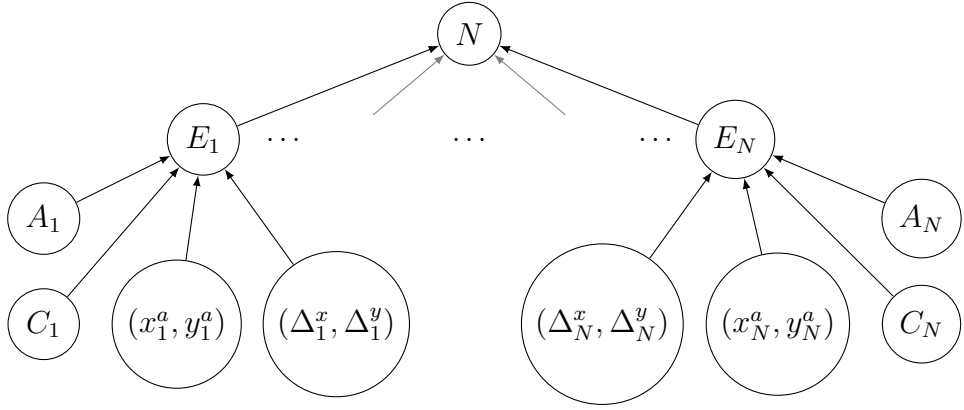

In this section we define our model to capture a team's chances,  
before discussing an approach to determine the composite nature of 
each individual chance. Each chance consists of an assist player, 
a player taking the chance (chance player), the spatial location from 
which the assist was made and the location of the chance. First, the 
number of chances a team has in a given period~($N$) is sampled using 
a Poisson model. Then for each chance ($E$), we draw an assist player 
($A$) and a chance player ($C$) from discrete distributions, with an 
assist location $(x^a,y^a)$ and the difference between the assist and 
chance locations $(\Delta^{x},\Delta^{y})$ being captured through 
Gaussian mixture models. A diagram of the model is given in 
Figure~\ref{pic-modelrep}. We begin by looking at the number of 
chances each team generates.

\subsection{A team's number of chances} \label{chances}

Consider the case where we have $K$ matches, numbered $k=1,\ldots,K$. 
We denote the set of teams in fixture $k$ as $T_k$, with $T_k^H$ and 
$T_k^A$ representing the home and away teams respectively. Explicitly, 
$T_k = \{T_k^H, T_k^A\}$. We take $P$ to be the set of all players who 
feature in the dataset, and $P^j\in P$ to be the subset of players 
who play for team~$j$.

\begin{figure}
\vspace{0.5cm}
\begin{center}
\begin{tikzpicture}[scale=7,>=latex]
     \draw[black]
         (0,0.0) rectangle (1.5,0.2); 
         
     \draw[black]
         (0,-0.05) -- (0,0.2);
     \draw[black]
         (0.25,-0.05) -- (0.25,0.2);
     \draw[black]
         (0.5,-0.05) -- (0.5,0.2);
     \draw[black]
         (0.75,-0.05) -- (0.75,0.2);
     \draw[black]
         (1.0,-0.05) -- (1.0,0.2);
     \draw[black]
         (1.25,-0.05) -- (1.25,0.2);
     \draw[black]
         (1.5,-0.05) -- (1.5,0.2);
          
     \node[draw,circle,fill=black,inner sep=0.5mm, label=below:{0}]
           (xo0) at (0,-0.05) {};         
     \node[draw,circle,fill=black,inner sep=0.5mm, label=below:{15}]
           (xo1) at (0.25,-0.05) {};             
     \node[draw,circle,fill=black,inner sep=0.5mm, label=below:{30}]
           (xo2) at (0.5,-0.05) {};             
     \node[draw,circle,fill=black,inner sep=0.5mm, label=below:{45}]
           (xo3) at (0.75,-0.05) {};    
     \node[draw,circle,fill=black,inner sep=0.5mm, label=below:{60}]
           (xo4) at (1.0,-0.05) {};               
     \node[draw,circle,fill=black,inner sep=0.5mm, label=below:{75}]
           (xo5) at (1.25,-0.05) {};               
     \node[draw,circle,fill=black,inner sep=0.5mm, label=below:{90}]
           (xo6) at (1.5,-0.05) {};               
              
     \node at (0.125, 0.1) {{$t_1$}};      
     \node at (0.375, 0.1) {{$t_2$}};      
     \node at (0.625, 0.1) {{$t_3$}};      
     \node at (0.875, 0.1) {{$t_4$}};      
     \node at (1.125, 0.1) {{$t_5$}};      
     \node at (1.375, 0.1) {{$t_6$}};  
     
     \node at (0.75, -0.25) {{Time (minutes)}};                                     
\end{tikzpicture}
\end{center}
\vspace{-0.5cm}
\caption{One possible way to split a fixture into blocks} \label{pic-matchblocks}
\end{figure}
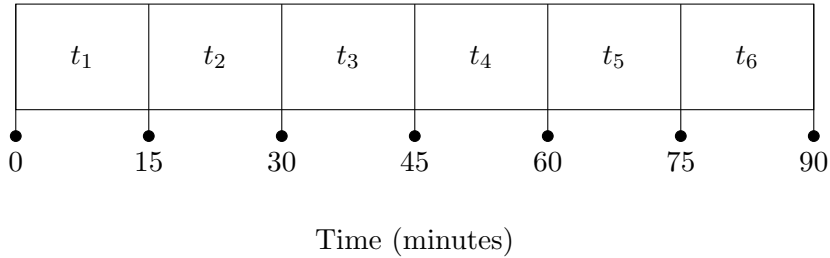

For simplicity we outline the model for a single fixture first. We 
split a fixture into blocks---one possibility being to split a 
fixture into 15 minute blocks, giving 6 blocks in total (see 
Figure~\ref{pic-matchblocks}). Of course the widths of these blocks is 
arbitrary, and could equally be set to be either a half of soccer (45 
minutes) or indeed every minute. After discussion with expert soccer 
analysts the authors feel that a block of 15 minutes provides sufficient 
granularity without introducing large levels of redundancy. Typically, 
a soccer match will have a small amount of extra time at the end of 
each half; throughout this paper, any chances which occur within these 
periods of extra time are included in either $t_3$ or $t_6$ (using the 
block structure illustrated in Figure~\ref{pic-matchblocks}).

Taking $N_{{t_r},k}^j$ to be the number of chances for team $j$, in 
match $k$, and block~$t_r$,~$r=~1,\ldots,6$, we have
\begin{equation} \label{posmod}
N_{{t_r},k}^j\sim Pois\left(\lambda_{{t_r},k}^j\right), 
\end{equation} 
where
\begin{equation} \label{lambda}
\lambda_{{t_r},k}^j = \exp\left\{\theta_{t_r}^j - \theta_{t_r}^{T_k\setminus j} + \left(\delta_{T_k^H,j}\right)\gamma_{t_r}  + \alpha G^j_{{t_r,k}} + \beta R^j_{{t_r,k}} \right\}.
\end{equation} 
A teams propensity to create chances is represented by $\theta_{t_r}^j$, 
$\theta_{t_r}^{T_k\setminus j}$ is the opposition's ability to 
create chances, $\gamma_{t_r}$ is a home effect for the corresponding 
block and $\delta_{a,b}$ is the Kronecker delta. The home effect 
reflects the (supposed) advantage the home team has over the away team.  
A home effect for the number of goals a team scores has been discussed 
by numerous authors, see for example \citep{dixon_1997, karlis_2003}. 
The current game state at the start of a block for a team is 
$G^j_{{t_r,k}}$, with $R^j_{{t_r,k}}$ being the red card state. For 
identifiability purposes, we follow \cite{karlis_2003} (amongst others) 
and impose the constraint that the $\theta_{t_r}^j$ must sum-to-zero, 
specifically 
\[
\sum_{i\in j} \theta_{t_r}^i = 0.
\]
The thinking behind this model construction is that if a team is creating 
chances, the other team cannot. Whilst this assumption is limiting by 
construction, given defensive tactics and other tangential aspects of 
play, it is the easiest (and possibly most meaningful) set-up derived 
from the data, which consists of attacking instances only. From 
\eqref{posmod} and \eqref{lambda}, the likelihood is given by
\begin{equation} 
L_N = \prod_{r=1}^6\prod_{k=1}^K\prod_{j\in T_k} \dfrac{\left(\lambda_{{t_r},k}^j \right)^{N_{{t_r},k}^j}\exp{\left(-\lambda_{{t_r},k}^j \right)}}{N_{{t_r},k}^j\,!}. \label{llike-pois}
\end{equation}

We note that it is possible to model the number of chances a team 
creates using an approach similar to the one implemented by 
\citep{dixon_1997, karlis_2003, baio_2010, whitaker_2017} (albeit for 
goals a team scores). However, we find little or no difference in the 
sum-of-squares, bias or empirical predictive distributions under the 
two set-ups. Thus, we proceed with the simpler model (in terms of the 
number of parameters) given by \eqref{posmod}--\eqref{llike-pois}.

\subsection{Chance composition} \label{chancecomposition}

Once the number of chances created by a team is determined by the above, 
we break $N_{{t_r},k}^j$ into separate events, $E_s$, where 
$s~=~1,\ldots,N_{{t_r},k}^j$ and 
\[
E=\left(E_1,\ldots,E_{N_{{t_r},k}^j}\right).
\] 
Each $E_s$ is a composition of the assist player ($A$), the chance 
player ($C$), the $(x,y)$-coordinates for the assist location 
$(x^a,y^a)$ and the difference between the assist and chance locations 
$(\Delta^{x},\Delta^{y})$, where 
\begin{align*}
\Delta^{x} & = x^c - x^a \\
\Delta^{y} & = y^c - y^a,
\end{align*}
with $(x^c,y^c)$ being the $(x,y)$-coordinates of the chance. By using 
the difference between the assist and chance locations we aim to model 
any dependence we may observe between the assist and chance locations. 
Explicitly $E_s=\left[A,C,x^a,y^a,\Delta^{x},\Delta^{y}\right]$. 

First, let us consider the task of determining the assist and chance 
player involved with each event. We make the assumption that a player 
cannot assist a player on an opposing team (such as assisting an own 
goal, by forcing the error), and neither can they take a chance created 
by a player from the opposition (for example, running onto a bad back 
pass). In the context of soccer these events are reasonably rare, and 
by implementing this assumption we can consider the players of one team 
to be independent from the players of another team. A player can 
switch teams part way through a season (in January) or at the end of a 
season by means of a transfer; however, we consider them to be a new 
player to be learned, as they may have different dynamics with their 
new team mates and possibly play in a different system, for example, 
playing in a new position to the one at their previous team. We model 
the probability of each assist player (and chance player) using a 
Multinoulli (or categorical) distribution. 

Let, $Z_{s,i,t_r}^a$ be a one-hot vector, with a 1 in position $i$, 
representing the assist player for event $s$, in a given block $t_r$, 
with $i\in P^j$. Denote the probability of each player 
making an assist for a given event by $\phi_{i,t_r}^a$, where 
\[
\sum_{i\in P^j}\phi_{i,t_r}^a = 1.
\] 
Setting $\phi^a_{t_r}$ to be the vector of $\phi_{i,t_r}^a$s, 
$\phi^a$ to be the vector of $\phi^a_{t_r}$s, $Z^a_{t_r}$ as the vector 
of $Z_{s,i,t_r}$s and $Z^a$, the vector of $Z^a_{t_r}$s, then 
\begin{equation}\label{multassist_a}
Z_{s,i,t_r}^a\sim \textrm{Multinoulli}\left(\phi^a_{t_r}\right),
\end{equation}
with 
\begin{equation}\label{zassist}
\pi\left(Z^a\big\vert\phi^a\right) = \prod_{r=1}^6\prod_{s=1}^{N_{{t_r},k}^j} \pi\left(Z_{s,i,t_r}^a\big\vert\phi^a_{t_r}\right).
\end{equation}
Similarly, for the chance player 
\begin{equation}\label{multassist_c}
Z_{s,i,t_r}^c\sim \textrm{Multinoulli}\left(\phi^c_{t_r}\right),
\end{equation}
where
\begin{equation}\label{zdelta}
\pi\left(Z^c\big\vert\phi^c\right) = \prod_{r=1}^6\prod_{s=1}^{N_{{t_r},k}^j} \pi\left(Z_{s,i,t_r}^c\big\vert\phi^c_{t_r}\right).
\end{equation}

Next, we consider the spatial locations, which we model using a 
mixture model. For a general discussion of mixture models 
we refer the reader to \cite{mclachlan_2004}. 
Given the nature of the spatial locations we implement a Gaussian 
mixture model, with $M$ components. Denote the weighting of the 
mixture components (for a given player $i$, in a given block $t_r$) 
by $\kappa_{i,t_r}^a$ and $\kappa_{i,t_r}^{\Delta}$ for the assist 
and $\Delta$ locations respectively, with 
$\kappa_{i,t_r}^*~=~(\kappa_{i,t_r,1}^*,\ldots,\kappa_{i,t_r,M}^*)$ and 
\[
\sum_{m=1}^M \kappa_{i,t_r,m}^*=1.
\]
Furthermore, let the observations for a given player, in a specific 
block, be $X_{i,t_r}^a$ and $X_{i,t_r}^{\Delta}$, with 
$X_{i,t_r}^*=(X_{i,t_r,1}^*,\ldots,X_{i,t_r,L_{i,t_r}^*}^*)$. Whence, 
the likelihood for the assist locations is 
\begin{equation}\label{llike-space_a}
L_a = \prod_{r=1}^6\prod_{i\in P} \prod_{l=1}^{L_{i,t_r}^a} \sum_{m=1}^M \kappa_{i, t_r, m}^a\times N\left\{ 
\begin{pmatrix}
x_{i,t_r,l}^a \\
y_{i,t_r,l}^a \end{pmatrix} ; 
\begin{pmatrix}
\mu^a_{x,m} \\
\mu^a_{y,m} \end{pmatrix}, \Sigma^a_m \right\},
\end{equation}
where $N(\cdot\,;\,m,V)$ denotes the multivariate Gaussian density with
mean $m$ and variance~$V$. 
Similarly
\begin{equation}\label{llike-space_d}
L_{\Delta} = \prod_{r=1}^6\prod_{i\in P} \prod_{l=1}^{L_{i,t_r}^{\Delta}} \sum_{m=1}^M \kappa_{i, t_r, m}^{\Delta}\times N\left\{ 
\begin{pmatrix}
x_{i,t_r,l}^{\Delta} \\
y_{i,t_r,l}^{\Delta} \end{pmatrix} ; 
\begin{pmatrix}
\mu^{\Delta}_{x,m} \\
\mu^{\Delta}_{y,m} \end{pmatrix}, \Sigma^{\Delta}_m \right\}.
\end{equation}

\begin{figure}[t!]
      \centering
      \includegraphics[scale=0.48]{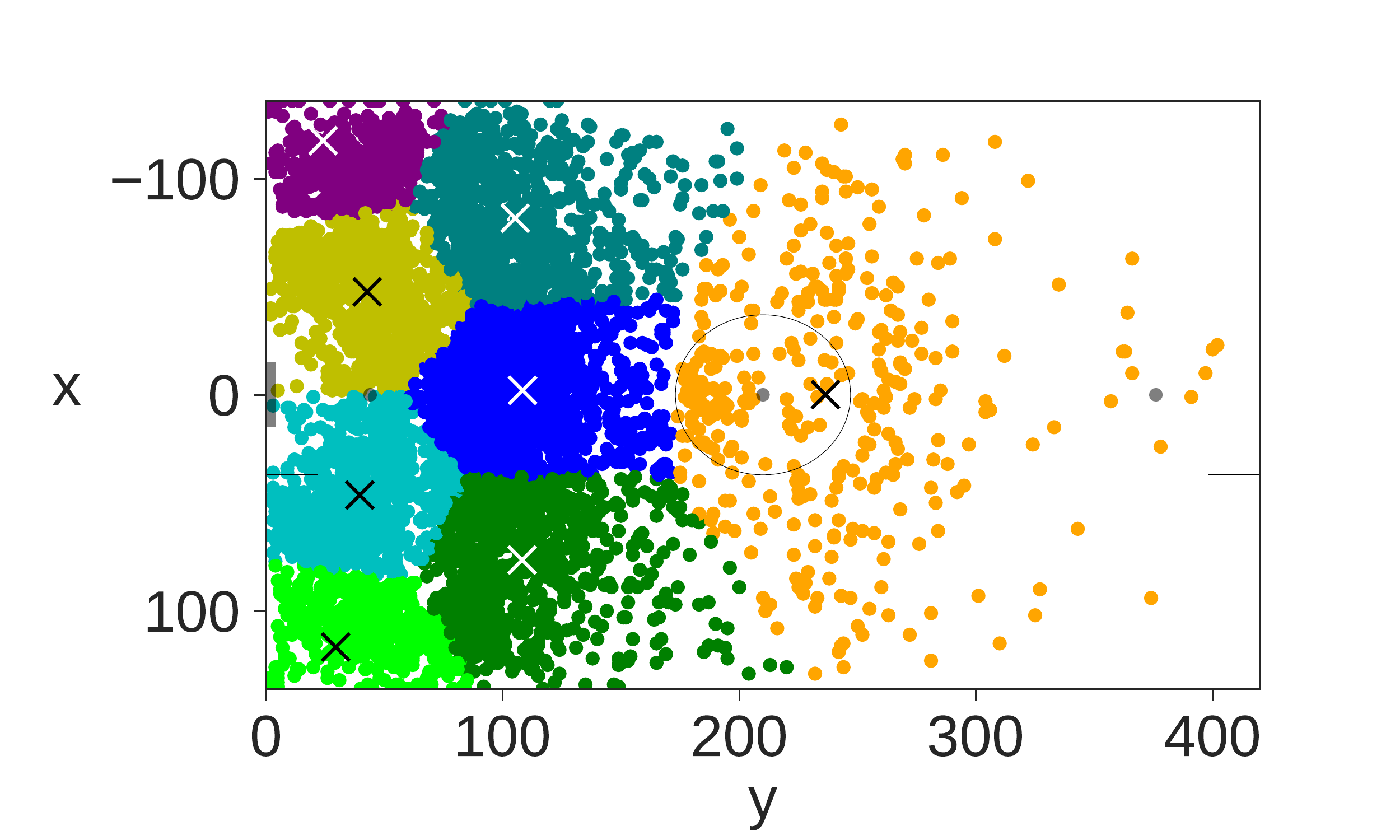}\vspace{0.05cm}
      \includegraphics[scale=0.48]{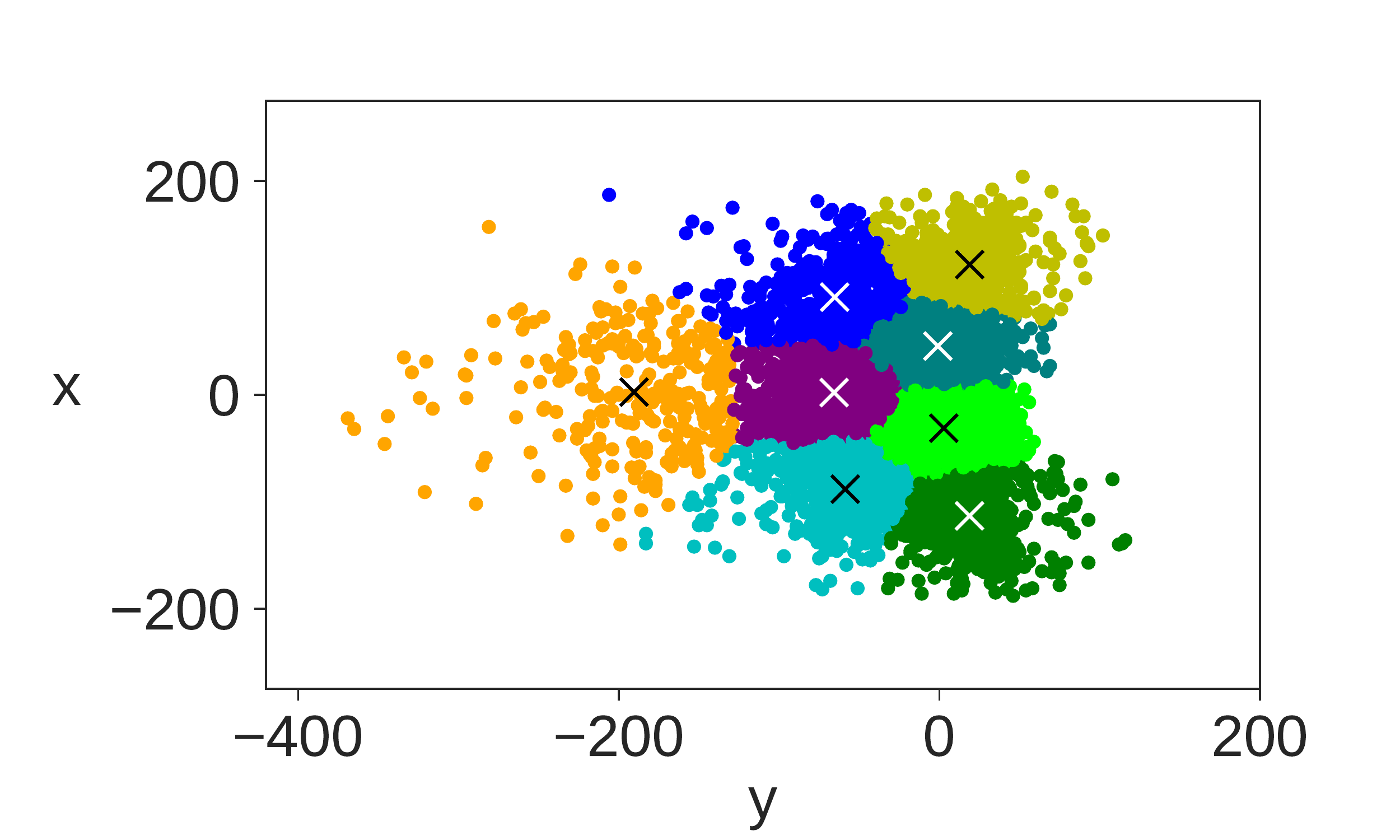}
      \caption{Cluster centroids (cross) under k-means with all data points classified by cluster assignment. \emph{Top} assist, \hbox{\emph{bottom}}~$\Delta$.} \label{clustfig}
\end{figure}

To simplify our approach we choose to predetermine the number of 
components which make up our mixture model. After discussion with 
expert soccer analysts we decided upon 8 components, whose locations 
we determine through k-means clustering. Thus, we set $\mu_m^a$, $m~=~1,\ldots,M$, 
to be the cluster centroids defined using all the observed assist 
locations (by all players), and $\mu_m^{\Delta}$, $m=1,\ldots,M$, using 
the $\Delta$ locations (deterministically constructed using the chance 
and assist locations). We leave $\Sigma^a_m$, $\Sigma^{\Delta}_m$, 
$m=1,\ldots,M$, as parameters to infer, rather than taking the 
variances of clusters per se. The locations of the cluster centroids 
are shown in Figure~\ref{clustfig} (indicated by a cross), where we 
also plot the data, colored according to cluster assignment (under 
k-means). To add some context to the cluster centroids, for the assist 
locations, the furthest right centroid $(0, 240)$ (own half, OH) is likely to 
represent a long ball forward for a player to run on to. For the leftmost 
column, the widest centroids $(x=-115, 115)$ (left corner [LC], right corner 
[RC]) are assists from corners or crosses into the box, whilst the middle 
two (left box [LB], right box [RB]) show cutbacks across goal and knock-downs. 
The middle column is slightly more ambiguous, although the center cross 
(center opposition half, CH) is most likely short through-ball assists, 
with the wider centroids being free-kicks and further crosses into the 
box (left opposition half [LH], right opposition half [RH]). The $\Delta$ 
centroids are the inverse of the assist centroids (in shape) and are 
simply the distance the ball traveled for the assist, for example, a 
larger magnitude of $x$ and a smaller magnitude of $y$ represents a 
cross into the box.

Having outlined the two components of our model, namely, the number of 
chances a team generates and the composition of these chances, we must 
consider the best way to fit the model, which is the subject of the 
next section.

\subsection{Bayesian inference}\label{bayes}

To estimate the parameters in the model we use a Bayesian inference 
approach. The joint posterior is given by
\begin{align}
& \mkern-40mu\pi\left(\theta, \alpha, \beta, \tau, \phi^a, \phi^c, \kappa^a, \kappa^{\Delta}, \Sigma^a, \Sigma^{\Delta}\big\vert N, Z^a, Z^c, x^a, y^a, \Delta^x, \Delta^y\right)   \nonumber \\
&\propto \pi\left(\alpha\right)\pi\left(\beta \right)\pi\left(\gamma \right)\pi\left(\tau \right)\pi\left(\theta\vert\tau \right)\pi\left(N\big\vert\theta,\alpha,\beta,\gamma \right) \nonumber \\
&\quad\times \pi\left(\phi^a\right)\pi\left(Z^a\big\vert\phi^a\right) \pi\left(\phi^c\right)\pi\left(Z^c\big\vert\phi^c\right)   \nonumber \\
&\qquad\times \pi\left(\kappa^a\right)\pi\left(\Sigma^{a}\right)\pi\left(x^a, y^a\big\vert \kappa^a, \mu^a, \Sigma^a, \phi^a\right) \nonumber \\
&\qquad\quad\times \pi\left(\kappa^{\Delta}\right)\pi\left(\Sigma^{\Delta}\right)\pi\left(\Delta^x, \Delta^y\big\vert \kappa^{\Delta}, \mu^{\Delta}, \Sigma^{\Delta}, \phi^c\right), \label{joint} 
\end{align}
where $\pi(N\vert\theta,\alpha,\beta,\gamma)$ follows \eqref{llike-pois}, 
$\pi(Z^a\vert\phi^a)$ is given by \eqref{zassist} and 
$\pi(Z^c\vert\phi^c)$ by \eqref{zdelta}, 
$\pi(x^a, y^a\vert \kappa^a, \mu^a,\Sigma^a, \phi^a)$ is governed by 
\eqref{llike-space_a} and 
$\pi(\Delta^x, \Delta^y\vert \kappa^{\Delta}, \mu^{\Delta}, \Sigma^{\Delta}, \phi^c)$ 
follows \eqref{llike-space_d}. Furthermore, $\pi(\theta\vert\tau)$ is 
the prior density ascribed to $\theta$, dependent upon $\tau$, which 
we take to follow a $N(0,\tau)$ distribution.

To fully specify the model, we implement the following priors \\
\begin{minipage}[b]{0.47\linewidth}
\begin{align}
\pi\left(\alpha\right)&\sim N\left(0,10^2\right), \nonumber\\ 
\pi\left(\gamma\right)&\sim N\left(0,10^2\right),\nonumber\\ 
\pi\left(\phi^a\right)&\sim\textrm{Dirichlet}\left(1_P\right), \nonumber\\
\pi\left(\kappa^a\right)&\sim\textrm{Dirichlet}\left(1_M\right), \nonumber\\
\pi\left(\Sigma^{a}\right)&\sim \mathcal{W}^{-1}\left(I_2, 2\right), \nonumber
\end{align}
\end{minipage} 
\begin{minipage}[b]{0.47\linewidth}
\begin{align}
\pi\left(\beta\right)&\sim N\left(0,10^2\right), \nonumber\\ 
\pi\left(\tau\right)&\sim \textrm{Gamma}(1, 0.01),\nonumber\\ 
\pi\left(\phi^c\right)&\sim\textrm{Dirichlet}\left(1_P\right), \nonumber\\
\pi\left(\kappa^{\Delta}\right)&\sim\textrm{Dirichlet}\left(1_M\right), \nonumber\\
\pi\left(\Sigma^{\Delta}\right)&\sim \mathcal{W}^{-1}\left(I_2, 2\right), \label{priors}
\end{align}
\end{minipage} \vspace{0.3cm}\\
where $1_q$ is a vector of 1s with length $q$, $I_q$ is the identity 
matrix with dimension $q$ and $\mathcal{W}^{-1}$ is the inverse Wishart 
distribution. By assuming $\phi^*$ follows a Dirichlet distribution 
\textit{a priori}, we are modeling the assist and chance players as a 
mixture of Multinomials, which is in line with techniques used in topic 
modeling, as part of a hierarchical Bayesian model. Where topic models 
(usually) capture the words for a particular topic, here, we determine 
the players for an assist or chance.

The form of \eqref{joint} admits a Gibbs sampling strategy with 
blocking, which we can extend to form five independent full 
conditionals for the number of chances, the assist player, the chance 
player, the location of the assist and the $\Delta$ location. Further 
blocking strategies that exploit the conditional dependencies between 
the model parameters and the data can also be used. To elucidate, the 
assist player, $\phi^a$, can be updated separately for each team. On 
top of this, all parameters can be updated separately for each block, 
$t_r$. We fit the model in \verb+Python+ using the package 
\verb+PyMC3+.

\section{Applications} \label{app}

Having outlined our approach to determine the number of chances a team 
will generate in a given fixture---accounting for the opposition's 
ability to create chances, the game and red card states and a home 
effect---plus a model for the composition of these chances, we wish to 
test the proposed methods in real world scenarios. Given the independence 
between the components which constitute the model we consider two 
applications. In the first we learn a team's ability to create chances 
and in the second we examine which players are involved, and where on 
the pitch these events occur. For both applications we use the data 
described in Section~\ref{data}, namely the 2016/2017 English Premier 
League. Throughout this section, to aid table/figure aesthetics, we 
refer to teams by the abbreviations given in Table~\ref{tab-teamabb}. 
We note that CHE won the league, with TOT, MCI and LIV getting UEFA 
Champions League places, therefore, we may expect these 4 teams to be 
the best. On the other hand, SUN, MID and HUL were relegated at the 
end of the season, meaning these 3 teams were perhaps the worst.

{
\begin{table} 
\footnotesize
\centering
\hspace{-1.1cm}
\begin{tabular}{ll|ll|ll|ll}
\hline
\multicolumn{8}{c}{Abbreviation\,/\,Team} \\
\hline
BOU &	AFC Bournemouth & EVE &	Everton & MUN &	Manchester United & SWA &	Swansea City \\
ARS &	Arsenal & HUL &	Hull City & MID &	Middlesbrough & TOT & Tottenham Hotspur \\
BUR &	Burnley & LEI &	Leicester City & SOU & Southampton & WAT &	Watford \\
CHE &	Chelsea & LIV &	Liverpool & STK &	Stoke City & WBA &	West Bromwich Albion \\
CRY & Crystal Palace & MCI &	Manchester City & SUN &	Sunderland & WHU &	West Ham United \\
\hline
\end{tabular}
\caption{2016/2017 English Premier League teams and abbreviations} \label{tab-teamabb}
\end{table} 
}

\subsection{Determining a team's chance ability} \label{teamchance-sec}

\begin{figure}
      \centering
      \includegraphics[scale=0.48]{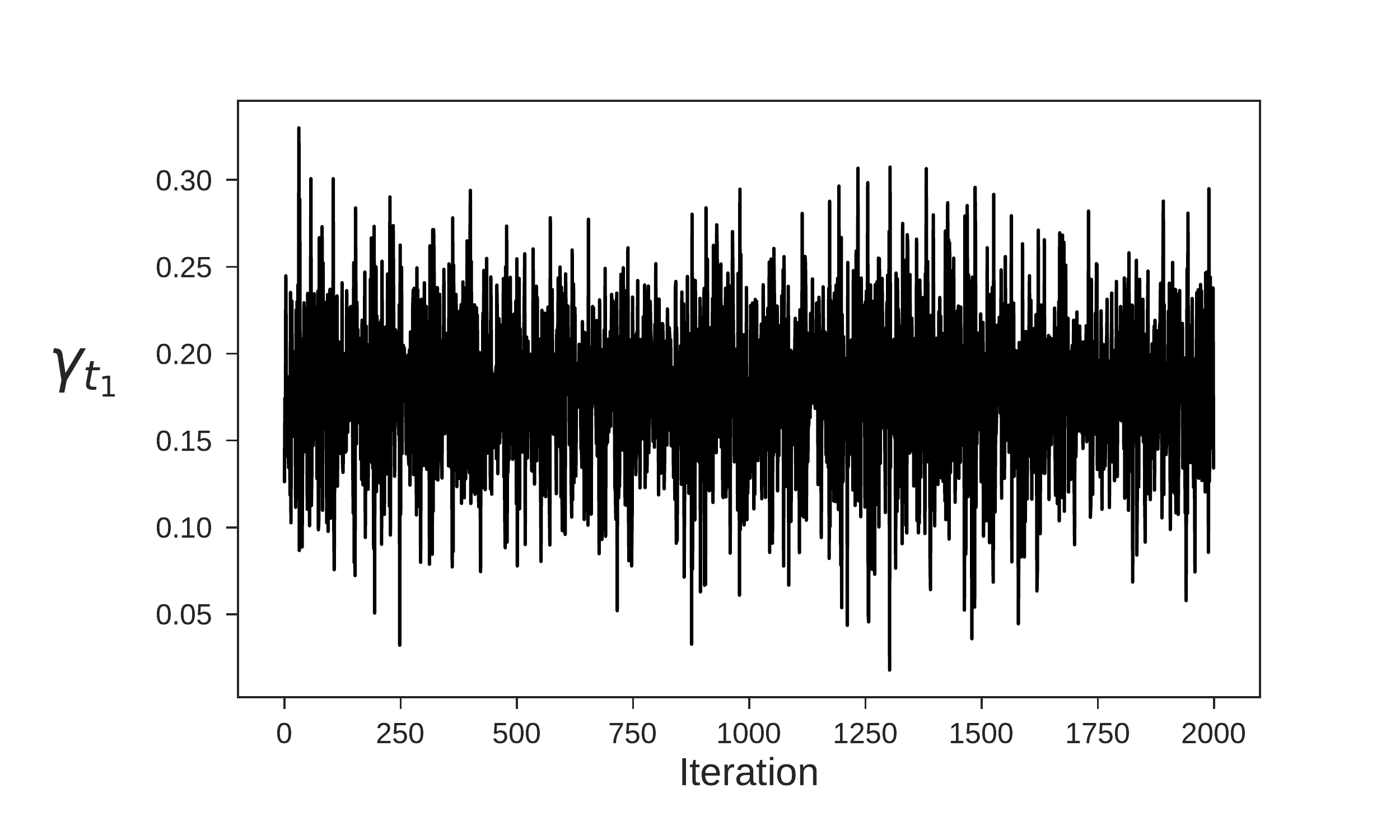}
      \caption{Trace plot for $\gamma_{t_1}$} \label{gammafig}
\end{figure}

We fit the model defined by \eqref{posmod}--\eqref{llike-pois}, using 
the priors specified in \eqref{priors}. We found little difference in 
results for alternative priors. We ran the model for 2000 iterations, 
after an initial burn-in of 100 iterations. A trace plot for $\gamma_{t_r}$ 
is given in Figure~\ref{gammafig}, where we see reasonable mixing (this 
trace plot is typical for all parameters in the model). 

\begin{table} 
\centering
\begin{tabular}{l|cccccc}
\hline
& \multicolumn{6}{c}{Block} \\
Team &	$t_1$ &	$t_2$ &	$t_3$ &	$t_4$ &	$t_5$ &	$t_6$ \\
\hline
BOU &	-0.043 &	-0.004 &	-0.211 &	-0.231 &	-0.003 &	-0.075 \\
ARS &	0.043 &	0.122 &	0.238 &	0.201 &	0.086 &	0.150\\
BUR &	-0.098 &	-0.174 &	-0.280 &	-0.166 &	-0.178 &	-0.291\\
CHE &	0.040 &	0.036 &	0.310 &	0.307 &	0.183 &	0.384\\
CRY & 	-0.0231 &	-0.079 &	-0.200 &	-0.106 &	-0.020 &	-0.126\\
EVE &	-0.024 &	0.030 &	0.233 &	0.015 &	0.069 &	0.284\\
HUL &	-0.143 &	-0.057 &	-0.304 &	-0.147 &	-0.183 &	-0.125\\
LEI &	-0.058 &	-0.080 &	-0.303 &	-0.121 &	-0.139 &	-0.174\\
LIV &	0.118 &	0.207 &	0.414 &	0.390 &	0.130 &	0.333\\
MCI &	0.201 &	0.401 &	0.375 &	0.268 &	0.249 &	0.465\\
MUN &	0.111 &	0.234 &	0.341 &	0.033 &	0.112 &	0.253\\
MID &	-0.065 &	-0.263 &	-0.255 &	-0.162 &	-0.208 &	-0.198\\
SOU &	0.092 &	0.075 &	0.090 &	0.132 &	0.020 &	0.091\\
STK &	-0.053 &	-0.182 &	-0.162 &	0.028 &	-0.081 &	-0.113\\
SUN &	-0.296 &	-0.080 &	-0.200 &	-0.156 &	-0.194 &	-0.531\\
SWA &	0.047 &	-0.139 &	-0.042 &	-0.093 &	-0.106 &	-0.236\\
TOT &	0.169 &	0.220 &	0.360 &	0.254 &	0.208 	& 0.332\\
WAT &	-0.095 &	-0.153 &	-0.189 &	-0.197 &	0.020 &	-0.211\\
WBA &	-0.001 &	-0.183 &	-0.160 &	-0.171 &	-0.015 &	-0.138\\
WHU &	0.077 &	0.071 &	-0.056 &	-0.076 &	0.051 &	-0.075\\
\hline
\end{tabular}
\caption{A team's mean ability to create chances, $\theta^j_{t_r}$, in the 2016/2017 English Premier League for each block} \label{tab-chances}
\end{table} 

The posterior 
means for a team's ability to create chances ($\theta^j_{t_r}$) over 
the entire 2016/2017 season are presented in Table~\ref{tab-chances}, 
for each of the 6 blocks. Those teams which we identified as possibly 
being ``better'' at creating 
chances, namely CHE, TOT, MCI and LIV, all have higher values in the 
table. Noticeably, they have higher values for blocks $t_5$ and $t_6$, 
when compared to other teams. This suggests they are able to find a way 
to win (by creating more chances) in the closing moments of a game (or 
a way to recover if they are losing), which is perhaps why they had a 
successful season. MCI have the highest value in $t_1$, $t_2$ and $t_6$,
meaning they started and finished games well. These values highlight 
Pep Guardiola's playing style, along with the quality of MCI's substitutes 
(they can replace good players with equally good players). CHE do not 
have as high values as some of the other top teams (even though they won 
the league), suggesting they did not create as many chances as other 
teams but they were more clinical with the ones they did create. 
Unsurprisingly, the teams who were relegated at the end of the season 
(SUN, MID, HUL) have some of the lowest values in the table. SUN have 
the worst ability to create chances in $t_1$ and $t_6$, with MID having 
a similar ability across all blocks, leading to them being the 2 lowest 
scoring teams in the league.

\begin{figure}
      \centering
      \includegraphics[scale=0.48]{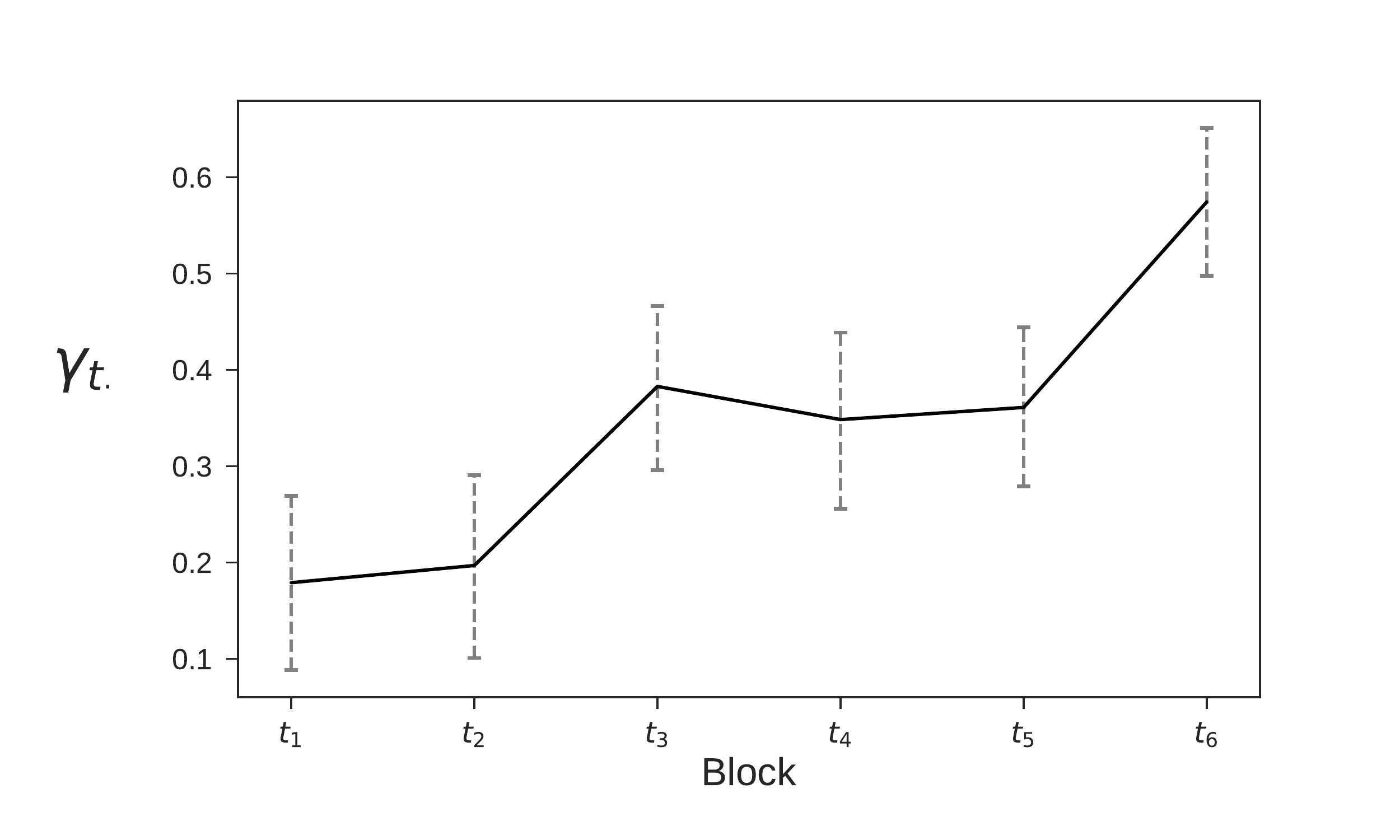}
      \caption{Mean home effect (solid line) and 95\% credible intervals (dotted line) in each block in the 2016/2017 English Premier League} \label{gammatimefig}
\end{figure} 

Figure~\ref{gammatimefig} shows the posterior mean for the home effect 
in each block over the entire 2016/2017 season, along with 95\% 
credible intervals. The credible intervals in each block are of 
near identical size, meaning we have similar levels of uncertainty 
surrounding all $\gamma_{t_r}$s. For all blocks we see a positive home 
effect, showing a team tends to create more chances at home than when 
playing away. This is in line with other findings concerning home 
effects. There is a rise in the home effect in $t_3$ (the end of the 
first half), this is possibly due to fan pressure to perform well. If 
a team is losing going into half time, fans want to see their team 
trying to get back into the game (by creating more chances), if they 
are drawing they want to try to gain an advantage, or if they are 
winning, they want to see them press home their advantage. This level 
of home effect carries into the second half $(t_4,t_5)$ before a 
similar rise is observed in $t_6$. The rise at the end of the game 
corresponds to a home team's desperateness to achieve a positive result 
(and please their fans). It is also possible that the home team is able 
to draw more energy from the crowd, and therefore out perform the away 
team. The trend seen in Figure~\ref{gammatimefig} compliments the 
findings of \cite{lucey_2013}, that a team will play more defensively 
away from home, with the suggestion that if an away team is winning or 
drawing in the final 15 minutes of a game ($t_6$), they will attempt 
to hold onto what they have (by defending more and creating less chances).

\begin{figure}
      \centering
      \includegraphics[scale=0.6]{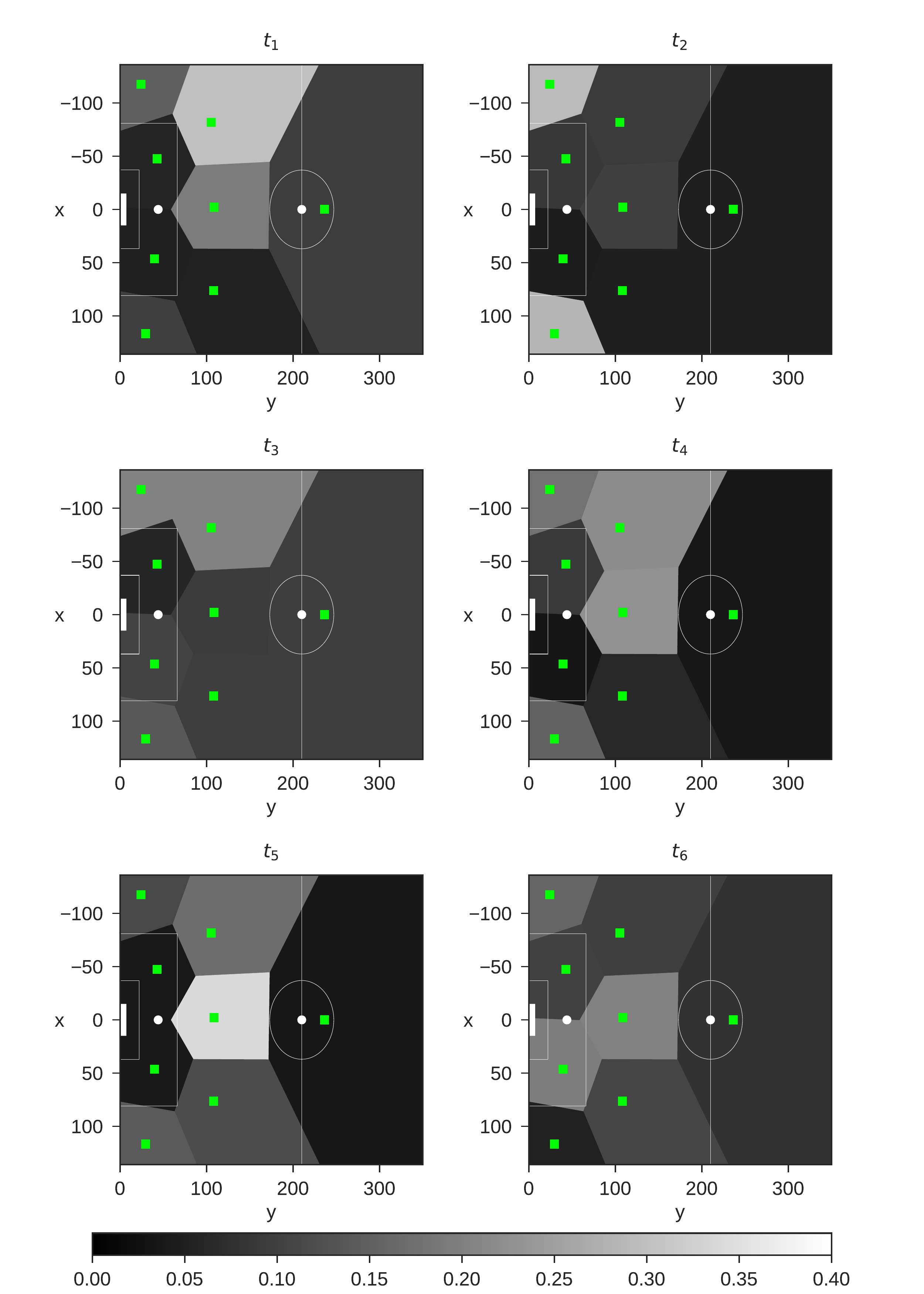}
      \caption{Eriksen assist locations for each block in the 2016/2017 English Premier League,  colored according to the weighting of each mixture component} \label{eriksenfig}
\end{figure} 

\begin{figure}
      \centering
      \includegraphics[scale=0.6]{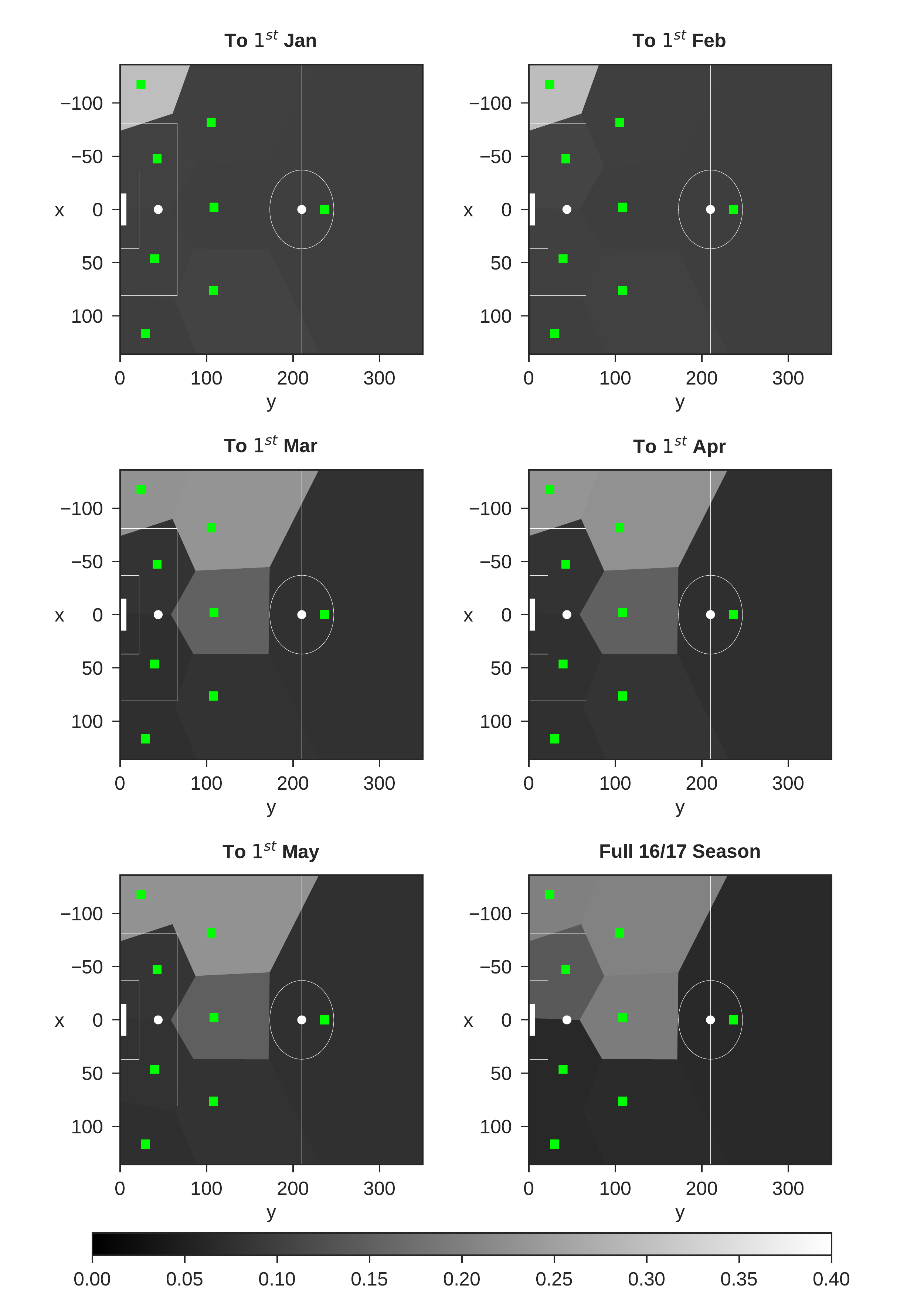}
      \caption{Mahrez assist locations in $t_5$ after different periods of time, colored  according to the weighting of each mixture component} \label{mahrezfig}
\end{figure} 

\subsection{Determining locations} \label{location-sec}

Having determined the number of chances a team will create, we now fit 
the model defined through \eqref{zassist} and 
\eqref{zdelta}--\eqref{llike-space_d} to capture the composition of 
these chances. Initially, we focus our attention on the assist and 
$\Delta$ locations. 

Christian Eriksen created the most chances in the 2016/2017 English 
Premier League. Figure~\ref{eriksenfig} illustrates the locations of 
these assists in each block through a Voronoi diagram, colored 
according to the weighting of each mixture component ($\kappa^a_{i,{t_r}}$). 
It clearly shows that Eriksen changes his style of play (or at least 
the location of his play, and possibly effectiveness) during different 
periods of the game, for example, the plots for $t_1$ and $t_2$. 

As we are implementing the model within the Bayesian paradigm, we can 
fit the model to a certain point in the season, before updating our 
beliefs once more data becomes available (more matches are played). To 
this end, we learn the model parameters using data up until 1/1/2017 
(roughly half the season), and then proceed to update our beliefs after 
each subsequent month. Voronoi diagrams for Riyad Mahrez's assists in 
$t_5$ after each of these months (along with the season as a whole) are 
shown in Figure~\ref{mahrezfig}. Mahrez was one of the stars for 
Leicester City when they won the league in 2015/2016, however he was 
not playing as well under manager Claudio Ranieri in 2016/2017 (our 
dataset); this is evidenced by the top row of plots where high weights 
are only assigned to the left corner. Ranieri was sacked in February 
and Craig Shakespeare became manager, who was seen to get Mahrez back 
playing somewhere near his best. The figure supports this, with the 
bottom 4 plots showing assists coming from more areas of the pitch, 
those being, the left-hand side, drifting to more central positions. 
This approach (through Figures~\ref{eriksenfig} and \ref{mahrezfig}) 
illustrates that we can model how a player plays throughout a game and 
over a season. Given we can update as more data becomes available, this 
allows us to capture when a player changes their style of play or when 
they start to become more/less important to a team.

\begin{figure}
      \centering
      \includegraphics[scale=0.6]{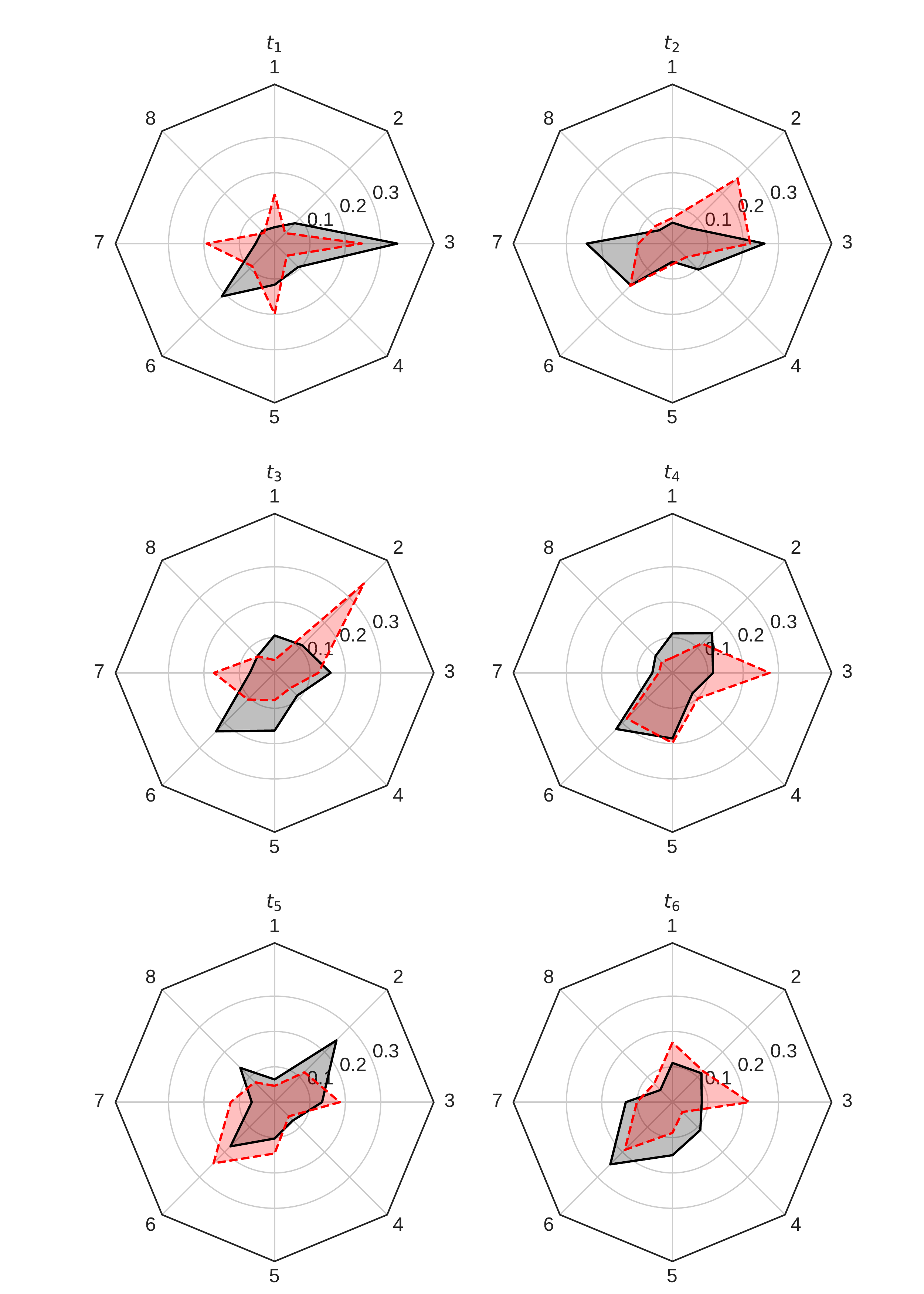}
      \caption{Radar plots of the mean $\kappa_{i,t_r}^{\Delta}$ for Kane (solid) and Ag\"{u}ero (dashed) for each block in the 2016/2017 English Premier League} \label{kaneaguerofig}
\end{figure} 

Integrating over the posterior uncertainty of the spatial locations 
gives the marginal posterior densities for $\kappa^{\Delta}$, from 
which we can ascertain differences in how certain players take chances. 
Radar plots of the mean $\kappa_{i,t_r}^{\Delta}$ (at each centroid) 
for Harry Kane (scored the most goals) and Sergio Ag\"{u}ero (had the 
most chances) are shown in Figure~\ref{kaneaguerofig}. For simplicity 
we number the centroids 1--8. The meaning of each centroid is subtle, 
and explanation is beyond the scope of this paper. However, it is clear 
from Figure~\ref{kaneaguerofig} that it is easy to visualize (and 
distinguish) between how certain players take chances, for instance, 
the differences in the shape of each player's radar plot.

\begin{figure}
      \centering
      \includegraphics[scale=0.48]{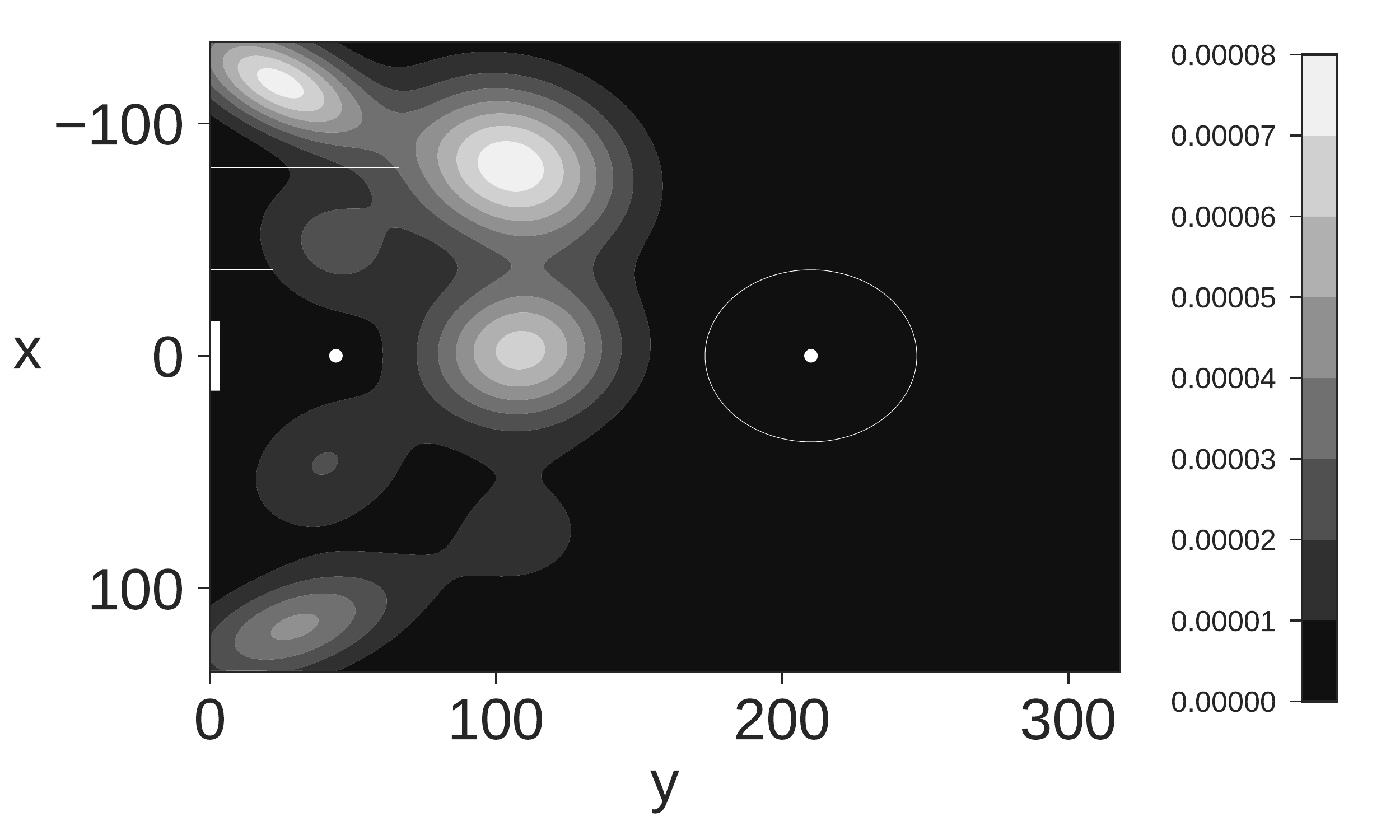}
      \caption{Eriksen assist locations under the Gaussian mixture model for $t_1$ in the 2016/2017 English Premier League} \label{eriksenGMMfig}
\end{figure}

By marginalizing over the mixture weights ($\kappa^*$), along with the 
uncertainty within the mixture components, we can construct a surface 
under the Gaussian mixture model for each player. One such surface is 
presented in Figure~\ref{eriksenGMMfig}. This is the surface for 
Christian Eriksen's assists in $t_1$ over the entire 2016/2017 English 
Premier League. Note, these surfaces can be constructed at any point 
in the season and updated once more data becomes available. From the 
figure it is easy to see where this player had most influence, and we 
observe a similar pattern to the one seen in the top left plot of 
Figure~\ref{eriksenfig}. Such plots are a useful way to convey 
information to a team, an application of which we consider below.

\subsubsection{Identifying a team's strengths and weaknesses}

During the 2016/2017 English Premier League many pundits questioned the 
ability of LIV's defense, highlighting a weakness on the left-hand side. 
Looking at the data, this criticism appears fair. Of the goals LIV 
concede, the assist leading to the goal is most likely to come from 
the left-hand side of the box (LB), with a $\Delta$ $(x,y)$-location of 
approximately (50,0), see Figure~\ref{clustfig} for cluster locations. 
Moreover, they are most likely to concede from these positions in 
blocks $t_3$ and $t_5$. Therefore, when approaching a game, LIV may 
want to know which of the opposition players are most likely to be 
involved in chances at these locations for each block, so that they can 
attempt to reduce their impact.

\begin{figure}
      \centering
      \includegraphics[scale=0.48]{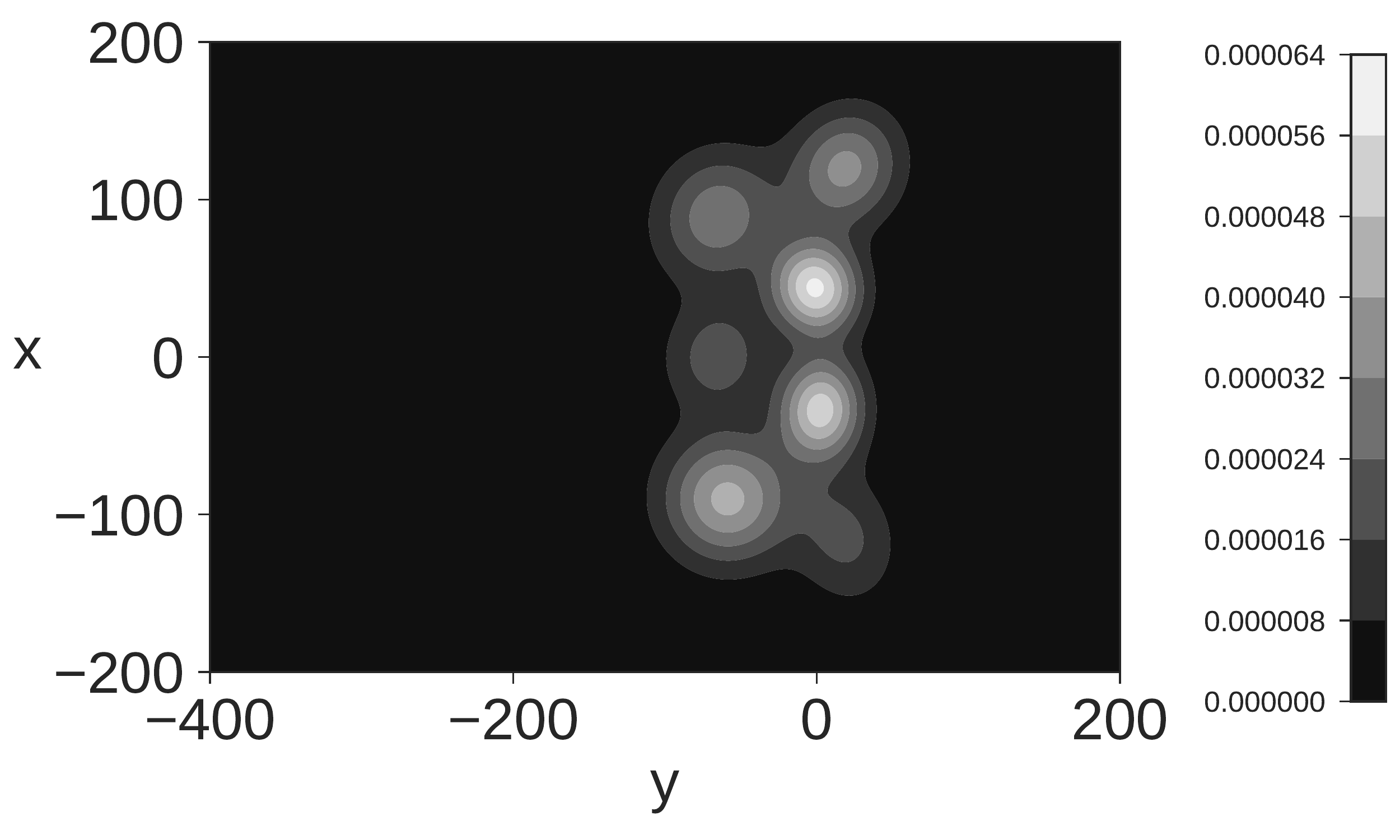}
      \caption{Benteke $\Delta$ locations under the Gaussian mixture model in $t_3$ using data to $22^{\textrm{nd}}$ April 2017} \label{GMMdeltafig}
\end{figure}

\begin{figure}[h!]
      \centering
      \includegraphics[scale=0.48]{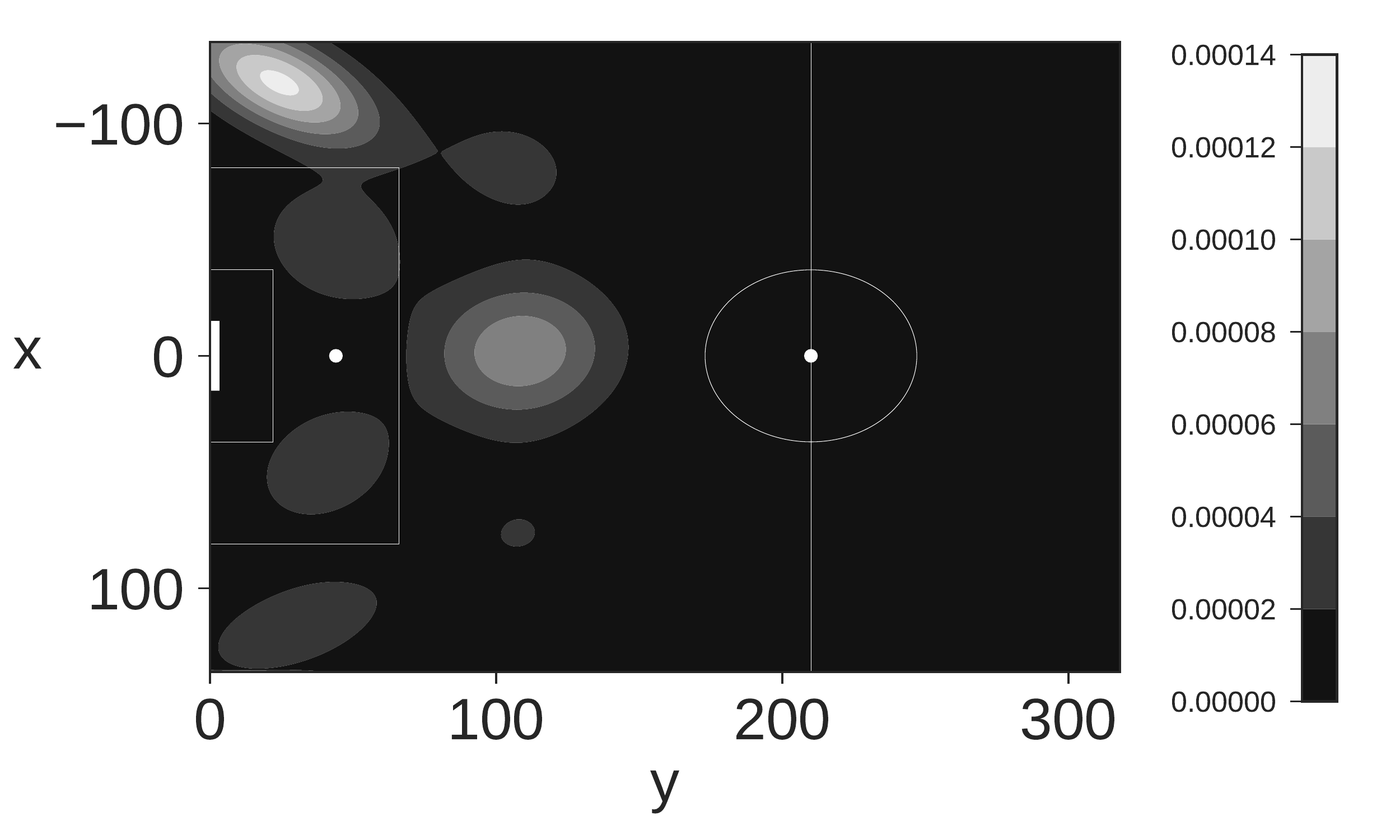}\vspace{0.1cm}
      \includegraphics[scale=0.48]{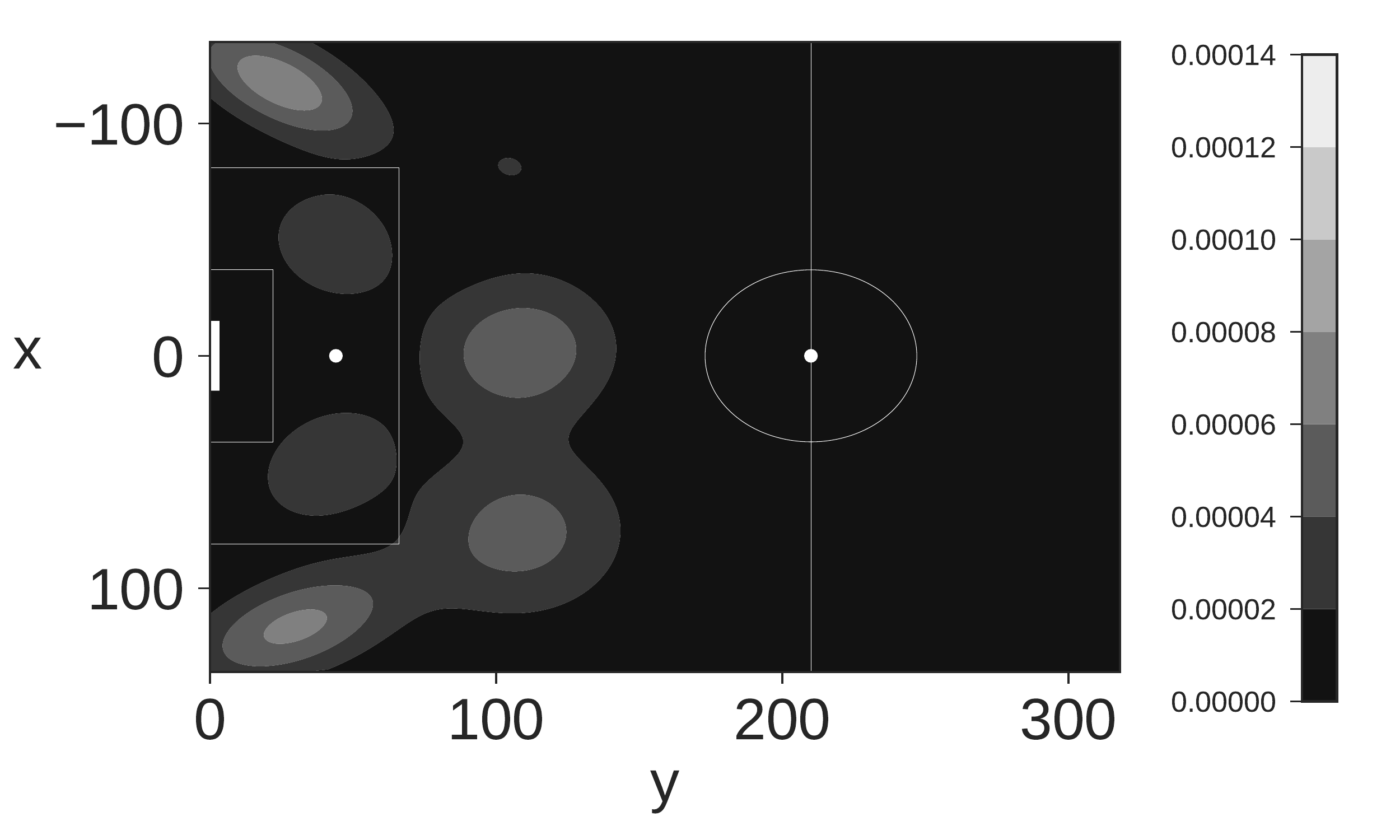}
      \caption{Assist locations under the Gaussian mixture model in $t_5$ using data to $22^{\textrm{nd}}$ April 2017. \emph{Top}~Cabaye, \hbox{\emph{bottom}}~Puncheon} \label{GMMassistfig}
\end{figure}

Let us consider the match LIV vs CRY (23/4/17)---CRY are a team who in 
recent years have caused LIV problems. We fit our model using all data 
available before the match is played. From the model, in both $t_3$ 
and $t_5$ we expect CRY to have 1 chance against LIV (in the match they 
had 2 chances in both $t_3$ and $t_5$). By integrating over $\phi^*$, 
$\kappa^*$, $\Sigma^*$ and by applying Bayes theorem we can calculate 
the probability of each player being involved in a chance, for each 
block, at LIV's weak locations. Christian Benteke is the most likely 
CRY player in $t_3$ to have a chance at the $\Delta$ location (with 
probability 0.166). Andros Townsend is the most likely in $t_5$, 
although there is little difference between the probability of Townsend 
and Benteke. Assists are likely to come from James McArthur in $t_3$, 
or Yohan Cabaye or Jason Puncheon in $t_5$ (with probabilities 0.134 
and 0.121 respectively). The $\Delta$ surface for Benteke in $t_3$ is 
shown in Figure~\ref{GMMdeltafig}, with the assist surfaces for Cabaye 
and Puncheon in $t_5$ given in Figure~\ref{GMMassistfig}. In both 
figures we see the highlighted ability of these players at the 
locations LIV are most susceptible. During the game LIV did not stop 
these players adequately enough, with Benteke scoring in both $t_3$ and 
$t_5$, Cabaye assisting in $t_3$ and Puncheon assisting in $t_5$.

\section{Discussion} \label{disc}

Within this paper we have provided a framework to determine the 
number of chances a team creates, along with the players and locations 
which make up a chance, in a Bayesian inference setting. Our approach 
is computationally efficient and utilizes the combination of a Poisson 
and Gaussian mixture model. We have shown in Section~\ref{app} that 
inferences under the model are reasonably accurate and have close ties 
to reality, along with implementable applications (of which we only 
illustrate a few). In contrast to 
previous work, we exploit coarser data to identify individual player 
contributions, rather than modeling the spatial dynamics of a team as 
a whole. 

There are a number of ways in which the current work can be extended. 
Firstly, smoothing techniques can be applied to $\phi^*$ and $\kappa^*$ 
so that the probabilities of players and mixture components vary 
smoothly over time (this was not implemented here for computational 
simplicity). Also, there is some dependence between the player 
assisting the chance and the player taking the chance. To elucidate, 
some players link up better with some players than others (often 
determined by the areas on the pitch in which they play). This 
dependence between $A$ and $C$ needs incorporating into the model; 
which could also allow some network analysis techniques to be 
implemented. Finally, as an extension to the applications for the 
proposed methods, an interesting area of future work is anomaly 
detection. This would allow us to detect a change in a player's level, 
for instance, becoming a starting player rather than a substitute could 
increase a player's contribution in the earlier blocks of a game 
($t_1$--$t_4$). Techniques discussed in \cite{heard_2010} could be used 
as inspiration for methods to detect these changes.

\newpage
\bibliographystyle{apalike}
\bibliography{references}

\end{document}